# Transition from anomalous kinetics towards Fickian diffusion for Si dissolution into amorphous Ge

Zoltán Balogh[a], Zoltán Erdélyi, Dezső L. Beke, Gábor A. Langer, and Attila Csik

Department of Solid State Physics, University of

Debrecen, P.O. Box.2, H-4010 Debrecen, Hungary

Hans-Gerd Boyen, Ulf Wiedwald, and Paul Ziemann

Institut für Festkörperphysik, Universität Ulm, D-89069 Ulm, Germany

Alain Portavoce[1], Christophe Girardeaux[1,2]

[1]Aix-Marseille Université, IM2NP, [2]CNRS, IM2NP (UMR 6242)

FST Campus de Saint-Jérôme, Avenue Escadrille Normandie Niemen - Case

142, F-13397 Marseille Cedex, France

Over the last years several experimental and theoretical studies of diffusion kinetics on the nanoscale have shown that the time evolution ($x \propto t^k$) differs from the classical Fickian law ($k_c$=0.5). However, all work was based on crystalline samples or models, so far. In this letter, we report on the diffusion kinetics of a thin amorphous-Si layer into amorphous-Ge to account for the rising importance of amorphous materials in nanodevices. Employing surface sensitive technics, the initial $k_c$ was found at 0.7±0.1. Moreover, after some monolayers of Si dissolved into the Ge, *kc* changes to the generally expected classical Fickian law with $k_c$=0.5.

---

[a] Corresponding author; address: Department of Solid State Physics, University of Debrecen, P.O. Box 2 H-4010 Debrecen, Hungary; Tel./Fax: +36 52 316 073; e-mail: bz0015@delfin.unideb.hu

66.30.Pa, 68.35.Fx, 83.85.St

Studies of diffusional movement of atoms through interfaces and of the resulting interface shift are nowadays of great interest not only from a fundamental research point of view (the search of diffusion laws valid on the nanoscale) but they are of practical importance as well. Recent progress in device technology allows to directly approach structures on the nanoscale, for which the knowledge of the corresponding diffusion laws is indispensable. Also related to this topic is the possible degradation of parts of electronic devices being often due to material transport through interfaces. Accordingly, in order to extend the lifetime of such devices a detailed understanding of the underlying diffusion processes is required.

During the past years several publications dealt with anomalous diffusion in different systems (e.g. Refs. [1,2,3,4,5]). Some of them showed that during the dissolution of a thin film into a single crystalline substrate, the shift of the interface was proportional to $t^{k_c}$, where t is the time. The value of $k_c$ kinetic exponent was not always equal to 0.5 (Fickian diffusion). Depending on two parameters $k_c$ was between 0.25 and 1 (e.g. Refs. [6,7,8]). These parameters were the diffusion asymmetry parameter $m'$, which measures the difference between the diffusivities in the film and the substrate in orders of magnitude, and the mixing energy $V$ determining the phase separation tendency.

It is important to note that there are numerous other examples of experimentally observed nonparabolic growth kinetics in the literature. However, in all of these cases the linear growth of phase(s) in solid state reactions was interpreted as a consequence of the reaction rate control at the interface(s)[9,10,11,12]. It is also worth noting here that strong diffusion asymmetry may lead not only to sharp interface shift with anomalous kinetics on the nanoscale but to sharpening of an initially diffuse interface even in completely miscible systems[13,14].

However, $k_c$ has not been determined either from computer simulations or experiments in amorphous systems in the above studies, although they play more and more important roles in applications.

With an ever increasing importance of amorphous materials in nanoscale applications due to their useful optical and electronic features[15,16], it is an important question whether anomalous diffusion can be experimentally observed in amorphous systems as well.

To answer this question, we studied the dissolution of a thin (~3nm for AES and ~1-4nm for XPS experiments) amorphous Si layer into a thick (~100nm) amorphous Ge layer by Auger Electron Spectroscopy (AES) and X-ray Photoelectron Spectroscopy (XPS). From isothermal heat treatments, the corresponding kinetic exponents were determined.

Germanium layers with a thickness of 100nm were prepared at room temperature by magnetron sputtering. The quality of the Ge surface was checked by atomic force microscopy revealing a mean roughness below 1nm for typical scan lengths of 1000μm. For the AES investigations the native Ge oxide layer was removed by ion bombardment followed by annealing the samples for 1 hour at 573K. The 3nm thick Si top layers were prepared by thermal evaporation. Absence of the Ge(52eV) peak indicated a complete coverage by the Si top layer. For the XPS experiments, the Ge specimens were dipped into 2% HF for 15 seconds. After cleaning in deionized water, the samples were stored in ethanol before transferring them into the UHV system. After additional annealing at 473K for two hours, the Ge surfaces were found to be completely free of any oxide. Subsequently, the 1 to 4nm thick Si layers were prepared by e-beam evaporation in situ. In this case, Ultraviolet Photoelectron Spectrocopy was applied for testing complete coverage by the Si layer, since low energy photoelectrons excited by UPS have very low inelastic mean free path (IMFP).

AES experiments were performed under ultra high vacuum, using an electron beam as primary excitation source. The beam energy was set to 2050eV and the beam current to 105μA during the whole experiment. Auger electrons were detected using a cylindrical mirror analyzer (CMA). The sample temperature was measured and regulated via a platinum-rhodium thermocouple placed on the surface of the sample. Both the Ge LMM transition at 1147eV and the Si LMM transition at 92eV were simultaneously recorded in-situ. The Si transition corresponds to an IMFP of 0.5nm while the Ge transition corresponds to an IMFP of 2.5nm (IMFP calculated from Ref. [17]).

For the XPS measurements we used an Al Kα source (non-monochromatized) with the electrons detected by a 100 MCD multichannel detector (SPECS). To monitor the Si/Ge interdiffusion, the Ge2p3/2 peak, the Ge3p doublet and Si2s peak were recorded. To account for maximum intensity (for faster recording) we used a pass energy of 100 eV. The temperature was set to 603 K and controlled by a pyrometer.

Let us first discuss the AES data. For this purpose, on the upper panel of Fig.1 two possible interdiffusion scenarios are given schematically. Scenario (a) assumes a strong diffusional asymmetry with Si being a fast diffusant within amorphous Ge but not vice versa. In this case, a sharp Si/Ge interface can be conserved. The second scenario (b) assumes a more symmetric diffusion leading to a broadening of the interface. Both assumptions result in a characteristic time dependence of the AES signals. Since the Si LMM Auger electrons have a very short inelastic mean free path in Si ($\lambda_{Si} \cong 0.5$nm, see e.g. Ref. 17) whereas the Ge LMM electrons exhibit a much larger value ($\lambda_{Ge} \cong 2.5$nm, see e.g. Ref. 17) in Si, one expects for the sharp interface scenario a temporal increase of only the Ge signal while the Si signal remains constant until the Si thickness approaches the limit of about $2\lambda_{Si}$ (Fig.1(a)). In the case of interface broadening, Ge reaches the surface quite fast resulting in an immediate decrease of the Si signal. This decrease would be much more pronounced than the increase of the Ge signal, since the AES is much more sensitive for Si than for Ge (Fig.1(b)). The corresponding experimental data are also presented in the lower panel of Fig.1, where the temporal evolution of the Si and Ge AES signals are given. Comparing this to Fig.1 (a) and Fig.1 (b) it is immediately clear that the experiment is in favor of a conserved sharp Si/Ge interface during the observation time of two hours. Detailed analysis of these data allows estimating a kinetic exponent of $k_c \approx 0.7$. Thus, the present data corroborate the idea that the Si/Ge interface remains abrupt during diffusion as had been concluded from earlier Auger depth profiles.[18]

In order to refine the above results, XPS measurements were performed allowing quantitative analysis. In evaluating the XPS spectra, the results of the AES measurements were used, i.e. it was assumed that interface remains sharp during the diffusion. Thus, from change of Ge2p/Ge3p signal ratio, the decrease of the silicon thickness due to its diffusion into the underlying Ge layer. In order to finally extract the kinetic coefficient $k_c$, the position of the interface (initial Si thickness minus the apparent Si thickness) is plotted vs. time on a log-log scale in Fig.2. Since a power function is linear in this representation, the slope is just equal to the exponent: $x \propto t^{k_c} \Rightarrow \ln(x) \propto k_c \ln(t)$.

As can be seen, in Fig.2, in order to fit the data points two straight lines are needed. This means, that $k_c$ is not a constant, but changes in time. In this way, an initial and final value of $k_c$ could be determined resulting in 0.7 for the initial and 0.53 for the final value. Thus, one concludes that the process starts anomalously (in non-Fickian way), but returns to the Fickian behaviour for longer times.

Table 1 summarizes the results obtained in the initial stage of the different measurements. It is clearly visible that the measured kinetic exponents agree quite well. Their values around 0.7 indicate that a large diffusional asymmetry causes the non-Fickian interface shift just like in crystalline systems. The column 'transition length' gives the approximate Si length dissolved into Ge for which an anomalous diffusion with exponent $\approx 0.7 \pm 0.1$ could be measured. Clearly, the diffusion anomaly is restricted to a few monolayer of Si only after the dissolution of more layers the diffusion kinetics returns to the classical Fickian law.

In this paper, the first experimental proof of non-Fickian diffusional kinetics in amorphous materials is provided. The present results together with those obtained for crystalline systems demonstrate that asymmetric diffusion causes deviation from the classical diffusion kinetics on the nanoscale independently of the material structure and diffusional mechanism. Therefore, anomalous diffusion due to diffusion asymmetry appears as a fundamental phenomenon in nature.

Moreover, in the present work not only the anomalous part of the diffusion process could be observed, but also the transition back to the classical Fickian behavior.

This work was supported by the OTKA Board of Hungary (Nos. K67969, K61253, D048594, IN 70181), the KPI (Nos. OMFB-01700/2006, OMFB-00376/2006). One of the authors (Z. Erdélyi) of this paper is a grantee of the `Bolyai János' scholarship. The work at Ulm was financially supported by DFG-SFB 569. Experimental support by Ch. Pfahler and A. Tschetschetkin (both Ulm University) is gratefully acknowledged.

**Table 1 Results from different measurements. The last column gives the dissolved Si layers for which the kinetic exponents have been evaluated**

| Sample. No | Method | Exponent | Transition Length |
|---|---|---|---|
| 1 | AES | 0.7 | >2.5ML |
| 2 | XPS | 0.65 | >1ML |
| 3 | XPS | 0.72 | 1.5ML |
| 4 | XPS | 0.70 | 3.5ML |
| 5 | XPS | 0.68 | 2.5ML |

**Fig. 1** Upper panel: Two possible scenarios for the Ge LMM and Si LMM AES signal intensity vs. time recorded at fixed temperature. a) The interface remains sharp, b) There is an interface broadening during intermixing Lower panel: Experimental curve at 623K. The Ge signal increases, whereas the Si remains constant, which indicates that the interface remains abrupt during the process. (see also the text)

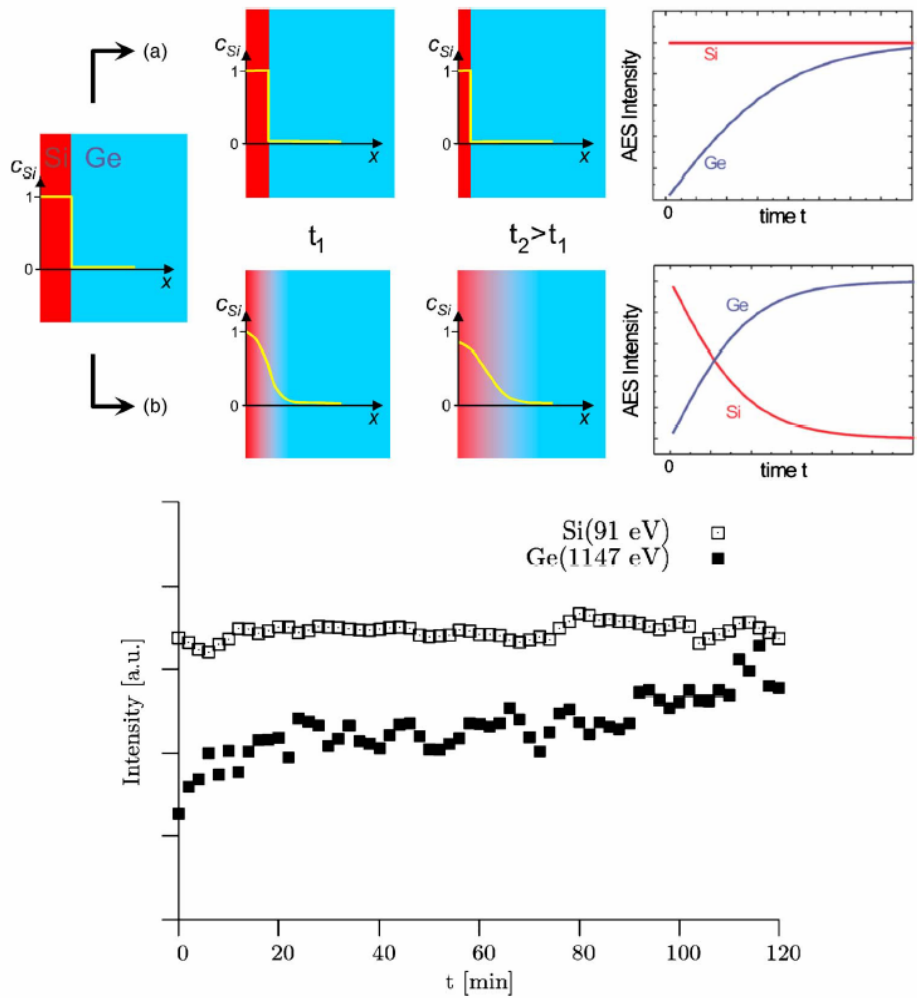

**Fig. 2** Interface shift (initial minus apparent thickness of the Si film) vs. time on log-log scale. The non-Fickian first part as well as the transition is clearly visible. (The dashed straight line is fitted to the first - anomalous - part of the data, whereas the solid one to the last - Fickian - part.)

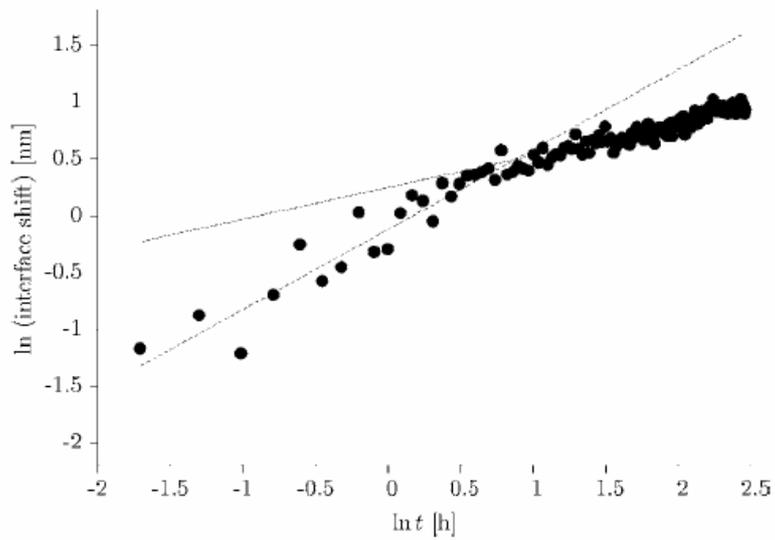